\documentclass{aa}  
\usepackage{graphicx}
\usepackage{natbib}
\usepackage{amsmath}
\usepackage{txfonts}
\bibpunct{(}{)}{;}{a}{}{,} % to follow the A&A style
\newcommand{\zs}{Z$_{\odot}$}
\newcommand{\ms}{M$_{\odot}$}

\newcommand{\mpc}{M$_{\odot}$/pc$^2$}
\newcommand{\dxpc}{dex kpc$^{-1}$}

\newcommand{\hmol}{H$_{\rm 2}$}
\newcommand{\hatm}{H${\rm I}$}

\begin{document}

   \title{Evolution of the Milky Way  with radial motions of stars and gas }

   \subtitle{II. The evolution of abundance  profiles from H to Ni}

   \author{M. Kubryk
          \inst{1},          
          N. Prantzos
          \inst{1}
          \and E. Athanassoula
          \inst{2}
          }

   \institute{Institut d'Astrophysique de Paris, UMR7095 CNRS, Univ. P. \& M. Curie, 98bis Bd. Arago, 75104 Paris, France\\
              \email{kubryk@iap.fr,prantzos@iap.fr}
         \and
             Aix Marseille Universit\'e, CNRS, LAM (Laboratoire d’Astrophysique de Marseille) UMR 7326, 13388, Marseille, France\\
             \email{lia@lam.fr}
                          }

   \date{Received ; accepted }

% \abstract{}{}{}{}{} 
% 5 {} token are mandatory
 
  \abstract
   {We study the role of radial motions of stars and gas on the evolution of abundance profiles in the Milky Way disk. }
   {We  investigate, in  a parametrized way,  the impact of  radial flows of gas and radial migration of stars induced mainly  by the Galactic bar and its iteraction with the spiral arms. }
   {We use a model with several new or up-dated ingredients (atomic and molecular gas phases, star formation depending on molecular gas, recent sets of metallicity-dependent stellar yields from H to Ni, observationally inferred SNIa rates), which reproduces well most global and local observables of the Milky Way.}
   {We obtain abundance profiles flattening both in the inner disk (because of radial flows) and in the outer disk (because of the adopted star formation law). The gas abundance profiles flatten with time, but the corresponding stellar profiles appear to be steeper for younger stars, because of radial migration.
We find a correlation between the stellar abundance profiles and O/Fe, which is a proxy for stellar age.    Our final abundance profiles are in overall agreement with observations, but  slightly steeper (by 0.01-0.02 \dxpc) for elements above S. We find an interesting "odd-even effect" in the behaviour of the abundance profiles (steeper slopes for odd elements) for all sets of stellar yields; however, this behaviour does not appear in observations, suggesting that the effect is, perhaps,  overestimated in current stellar nucleosynthesis calculations. }
   {}


   \maketitle

\section{Introduction}

The abundance profiles of chemical elements constitute one of the key properties of galactic disks. They depend on the past history of the disk and the various physical effects that affected it: star formation, infall and outflows, radial flows of gas, radial motions of stars and tidal interactions or mergers with other galaxies. Most semi-analytical studies of abundance profiles  were performed in the framework of the so-called "independent-ring" model, where the galactic disk is simulated as an ensemble of independently evolving annuli 
(e.g. \citet{Guesten1982,Matteucci1989,Ferrini1994,PranAub1995,Chiappini1997,BP99,Hou2000,Prantzos2000}) 
and concerned the MW disk, for which a large number of other constraints,
 both local and global are available. Those studies focused mainly on the interplay between the local star formation and infall rates, or the impact of variable stellar
  IMF. They also revealed the key issue of the evolution of the abundance profile, some studies supporting a flattening of it with time (e.g.
   \citet{Ferrini1994,PranAub1995,Hou2000} while others concluded the opposite (e.g. \citet{Tosi1988,Chiappini1997}).

Pioneering work of \cite{Tinsley1978} and \cite{Mayor1981} noticed the potential importance of radial gaseous flows for the chemical evolution of galactic disks. \cite{Lacey1985} presented a systematic investigation of the causes of such flows, and explored with parametrized calculations the impact of such effects on the chemical evolution of the Galaxy.  Further parametrized investigations  with simple 1D models of disk evolution are made in e.g. \cite{Tosi1988,Clarke1989,Sommer1990,Goetz1992,Chamcham1994,Edmunds1995,Portinari2000} and more recently in \cite{Spitoni2011,Bilitewski2012,Mott2013,Cavichia2014}. They have various motivations (mostly to fit the abundance profiles, but also gas and star profiles) and they are generally applied to the study of the MW disk. As expected, results are not conclusive, because they depend not only on the parametrization of the unknown inflow velocity patterns, but also on the other unknown (and parameterized) ingredients of the models, especially the adopted SFR and infall profiles as  functions of time.  
 Among the alleged causes of radial inflows, the impact of a galactic bar is well established, both from simulations and from observations. Numerical simulations \citep{Athanassoula1992,Friedli1993,Shlosman1993} showed  that the presence of a non-axisymmetric potential from a bar can drive important amounts of gas inwards of corotation (CR)  fuelling star formation in the galactic nucleus, while at the same time gas is pushed outwards outside corotation. In a disk galaxy, this radial flow mixes gas of metal-poor regions into metal-rich ones (and vice-versa) and may flatten the abundance profile  (e.g. \citet{Friedli1994,Zaritsky1994,Martin1994,Dutil1999}), although \cite{Sanchez2012} find little difference in that respect between barred and non-barred disks.  The study of \cite{KPA2013} suggests that bars may
be changing the chemical abundance profile inside the corotation
radius but they  have only a small impact outside it, while \cite{Martel2013} find a rather complex situation of continuous exchange of gas and metals between the bar and the central region of the disk.

The investigation of radial motions of stars - due to inhomogeneities of the galactic gravitational potential - on the chemical evolution of disks, has a more recent history. The role of the  bar has been studied to some extent with N-body+SPH codes by \cite{Friedli1993} and \cite{Friedli1994}. Observations in the 90ies revealed that the  MW does have a bar \citep{Blitz1991}, but its  size and age are not well known yet.  \cite{SellwoodBinney2002}  showed  that,  in the presence of recurring transient spirals, stars in a galactic disk could 
undergo important radial displacements: stars 
found at corotation with a spiral arm may  be scattered to different galactocentric radii (inwards or outwards),  a process which preserves
overall angular momentum distribution and does not contribute to the radial heating of the stellar disk.
Using a simple model, they showed how this process can increase the dispersion in the local metallicity vs age relation, well above the amount due to the epicyclic motion.
%This development paved the way for a large number of theoretical studies on radial migration, both with  N-body codes (e.g. \citealt{Roskar2008,Sanchez2009,Martinez2009,Sales2009,Roskar2010,minchevfamaey2010,Minchev2011,Brunetti2011,Minchev2012a, Grand2012,Baba2013,Bird2013,DiMatteo2013,KPA2013,Grand2014}) and with semi-analytical models \citep{Lepine2003, Prantzos09, SB2009, Minchev2013, Wang2013, Minchev2014}. Because of the difficulty to produce realistic MW-like disks, the former class of models focused mostly on generic properties of radial migration (origins of it and impact on some observables), while the latter focused exclusively on the properties of the MW.
\cite{minchevfamaey2010} suggested that resonance overlap of the bar and spiral structure \citep{Sygnet1988} produces a   more efficient redistribution of angular momentum in the disk. This bar-spiral coupling was studied in detail with N-body simulations by \cite{Brunetti2011} who
 found that  radial migration can be assimilated to a diffusion process, albeit with time- and position-dependent diffusion coefficients. That idea was confirmed by the analysis of N-body+SPH simulations of a disk galaxy by \cite{KPA2013} who showed that radial migration moves around not only "passive" tracers of chemical evolution (i.e. long-lived stars, keeping on the surfaces the chemical composition of the gas at the time and place of their birth), but also "active" agents of chemical evolution, i.e. long-lived nucleosynthesis sources (mainly SNIa producing Fe and $\sim$1.5 \ms \ stars producing s-process elements).

The implications of radial migration for the chemical evolution  of MW-type disks
were studied  with N-body codes by  \cite{Roskar2008} , who found that the stellar abundance profiles flatten with stellar age, even if the gaseous abundance profiles were steeper in the past. 
\cite{SB2009} introduced a parametrised prescription of
radial migration (distinguishing epicyclic motions 
from migration  due to transient spirals) in a semi-analytical chemical evolution code. They suggested that radial mixing could explain  not only local observables (e.g. the dispersion in the age-metallicity relation) but also the formation of the Galaxy's thick disk, by bringing to the solar neighborhood a kinematically "hot" stellar population from the inner disk. 
That possibility was subsequently investigated
with N-body models, but controversial results are obtained up to now:
while \cite{Loebman2011} find that secular processes (i.e. radial migration) are sufficient to explain the kinematic properties of the local thick disk, \cite{Minchev2012b} find this mechanism insufficient  and suggest that an external agent (e.g. early mergers) is required for that.

Following the pioneering work of \cite{SB2009}, the properties of the MW disk were studied in detail with semi-analytical models accounting for radial migration by
\cite{Minchev2013} and \cite{Kubryk2014}. The three models differ in several ways:
\cite{SB2009} use a toy-model of star transfer between adjacent radial zones (with coefficients tuned to reproduce properties of the local disk), whereas the other two are inspired by the results of N-body simulations (but they adopt different implementation techniques of those results). Radial gaseous flows are included in \cite{SB2009} and \cite{Kubryk2014}, but not in \cite{Minchev2013};  however, in \cite{SB2009} the radial flows concern mainly the outer disk, where in \cite{Kubryk2014} they concern the inner disk, since that work simulates the action of a bar. The dimension vertical to the galactic plane is considered in  \cite{SB2009} and \cite{Minchev2013}, but not in  \cite{Kubryk2014}. Finally, the star formation and radial infall laws are different in the three works. All models consider explicitly Fe production by SNIa (albeit with different prescriptions for the SNIa rate) and the finite lifetimes of stars.

Despite those differences, all three models find good agreement with the main observables of the MW, both locally (dispersion in age-metallicity relation, metallicity distribution, the characteristic "two-branch" behaviour between thick and thin disk in the O/Fe vs Fe/H plane) and globally (stellar and abundance profiles). This agreement suggests that, despite their sophistication, such models still involve too many parameters and suffer from degeneracy problems. We note
here the  difference in the final abundance gradient of Fe/H between \cite{SB2009} and \cite{Minchev2014}, who find slopes of the corresponding exponential profiles of -0.1 \dxpc and -0.06 \dxpc, respectively. Recent observations of statistically significant samples of Cepheids are consistent with the latter value, as we shall discuss in Sec. \ref{sec:abundance-evol}.

In this work, we study the evolution of abundance profiles of all elements from H to Ni  in the MW,  using the model  presented in  Kubryk et al. (2014, hereafter KPA2014).
The plan of the paper is as follows: The main  ingredients of the model are briefly presented  in Sec. \ref{sec:Model}, where we also discuss some of the results concerning the impact of radial migration on the disk properties.  We illustrate that impact by comparing a model with radial migration to one without it. In Sec. \ref{sec:abundance-evol} we discuss in some detail the profiles of the most important metals, namely O (Sec. \ref{sub:O-profiles})  and Fe 
(Sec. \ref{sub:Fe-profiles}) and we compare them to a large number of recent observations from various metallicity tracers. The impact of radial migration on 
the evolution of the abundance profiles is discussed in Sec. \ref{subsec:Evolution_profiles}, where we compare our results to those of a similar study \citep{Minchev2014} and to a compilation of extragalactic observations by \cite{Jones2013}. In Sec. \ref{sub:OFe} we adress the issue of the O/Fe ratio; in view of the small dispersion displayed by that ratio as a function of time, it can be used as a robust proxy for stellar age in studies of the evolution of the abundance profiles in the disk. In Sec. \ref{sub:Other} we present our results for all elements from H to Ni and we compare them to several sets of observational data. We find good overall agreement with observations, but a systematically larger (in absolute value) slope of the abundance profiles for the Fe-peak elements compared to observations. We also reveal - and we draw attention to -  interesting differences between the results obtained with different sets of stellar yields, as well as a 
manifestation of the "odd-even" effect of nucleosynthesis, which does not appear, however, in the observational data.
A summary of the results is presented in Sec. \ref{sec:Summary}.

\section{The model}
\label{sec:Model}

\begin{figure}
\begin{center}
\includegraphics[width=0.49\textwidth]{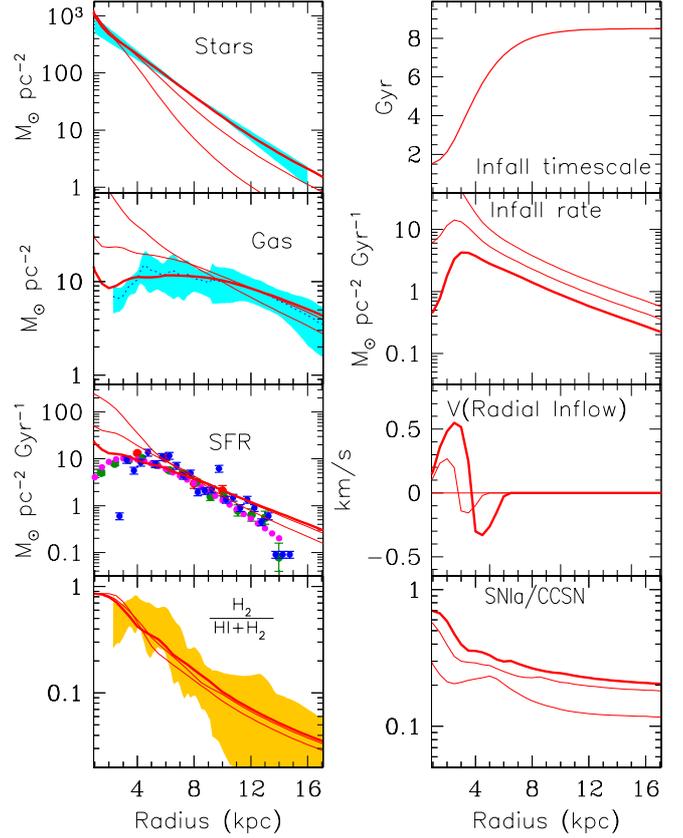}
\caption[]{{\it Left } (from top to bottom): Model profiles of stars, gas, SFR and molecular gas fraction $f_{Mol}$ at 4, 8 ({\it thin} curves) and 12 Gyr ({\it thick} curve). The curve at 12 Gyr is compared to observational data for the present-day profiles of the correspoding quantities ({\it shaded aereas} for the stars, gas and $f_{Mol}$ and points with error bars for the SFR); data sources are provided in KPA14. {\it Right} (from top to bottom): Infall timescales, infall rates, velocity profiles of radial inflow (positive values towards the Galactic center), ratio of SNIa/CCSN ; in all panels but  the top one, the three curves correspond to 4, 8 and 12 Gyr (thickest curve). 
}
\label{Fig:ModelGen1}
\end{center}
\end{figure}

The model presented in KPA14 for the evolution of the MW disk, involves radial motions of both gas and stars. The MW disk is built gradually  by infall of primordial gas \footnote{The composition of the infall is equally important when it comes to discuss the evolution of abundances and abundance ratios in the MW disk. Observations are of little help at present: they generally find low metallicities for gas clouds {\it presently} falling to the MW disk ($\sim$0.1 \zs, e.g. \citealt{Wakker1999}), but they provide no information on the past metallicity of such clouds or on their abundance ratios. Here we adopt the simplest possible assumption, namely that the infalling gas has always primordial composition. This assumption hardly affects the results for the chemical evolution of the disk, but it allows for the existence of  disk stars with metallicities lower than [Fe/H]=-1 (see \citet{Bensby2013a} and references therein).} in the potential well of a "typical" dark matter halo of final mass 10$^{12}$ \ms, the evolution of which is extracted from numerical simulations  (from 
\citet{Li2007}).  The infall rate is a parametrized function, its timescale increasing monotonically with galactocentric radius and ranging from 1 Gyr at 1 kpc to 7 Gyr at 7 kpc and slightly increasing further outwards (see Fig. \ref{Fig:ModelGen1}). Star formation depends on the local surface density of molecular gas, which is calculated by the semi-empirical prescriptions of \cite{BlitzRos06}:  it depends on a combination  of the stellar surface density profile (steeply decreasing with radius) and the gas surface density profile (essentially flat today).  This allows us to use the final profiles of atomic and molecular gas as supplementary constraints to the model (see Fig. 6 in KPA14). The adopted prescription produces a steep profile of \hmol \ (as observed in the Galaxy) and, thereoff, a steep SFR profile in the inner disk during most of the Galactic evolution;  this impacts directly on the corresponding abundance profiles, as we discuss below. For the radial flows of gas, we consider only the case of a MW-like bar operating for the last 6 Gyr and driving gas inwards and outwards of corotation. The  radial velocity profile of the gas flow induced by the bar is similar to the one adopted in \cite{Portinari2000} (their Case B for the bar), but not exactly the same: \cite{Portinari2000} consider additional radial flows inwards, acting all over the disk, while we limit ourselves to the case of the bar alone. The adopted radial velocity profile is given in Fig. \ref{Fig:ModelGen1}.  

We considered separately the epicyclic motion of stars (blurring) from the true variation of their guiding radius (churning), as in \cite{SB2009}. 
For the former, we developed an analytic formalism
based on the epicyclic approximation; for the latter, we adopted a parametrised description, using time- and radius- dependent diffusion coefficients, extracted from the N-body+SPH simulation of KPA13, which concerns a disk galaxy with a strong bar; as discussed in KPA2014, we adapted those transfer coefficients taking into account the smaller size of the MW bar. For the chemical evolution, we adopted recent sets of metallicity-dependent yields from \cite{Nomoto2013}, providing a homogeneous and fine grid of data, well adapted to the case of the MW disk. For comparison purposes, we also used older  yields from \cite{WW95} and \cite{CL2004}. We adopted the stellar IMF of \cite{Kroupa02} with a slope of X=1.7 (the Scalo slope) for the high masses. For the rate of SNIa, we adopted the empirical law of $R_{SNIa} \ \propto t^{-1.1}$, after the observed delayed time distribution of those objects in external galaxies (see e.g.  \cite{Maoz2012} and references therein).   In contrast with usual practice in studies of galactic chemical evolution, we adopted the formalism of Single Particle Population (SSP), which is the only one applicable to the case of radial migration, since it allows one to consider the radial displacements of nucleosynthesis sources and in particular of SNIa (see KPA13 for a discussion of that effect). Finally, we adopted a large and diverse set of recent observational data to constrain our model.

Our model reproduces well the present day values of most of the main global observables of the MW bulge (assumed to correspond to radii $r<$2 kpc) and disk ($r>$2 kpc): present-day masses of stars, atomic and molecular gas, star formation rates as well as core collapse supernova (CCSN) and SNIa rates (see Fig. 2 in KPA14). The corresponding  radial profiles of all those quantities (azimuthally averaged) are also reproduced in a satisfactory way (Fig. 6 in KPA14). The azimuthally averaged radial velocity of gas inflow in the bar region is constrained to be less than a few tenths of km/s in the framework of that model. The local properties of the MW disk, i.e. metallicity distribution and age-metallicity relation, are also well reproduced. In particular, following \cite{SellwoodBinney2002},   we showed how radial migration can be  constrained by the observed dispersion in the age-metallicity relation. . We emphasize, however, that the observational samples that we used - from  \cite{Bensby2014} for the age-metallicity relation and from \cite{Adibekyan2011} and \cite{Bensby2014} for the metallicity distribution -   have various selection biases (kinematic, limited by magnitude or volume), which have not been applied to our results: the model predictions for the solar neighborhood concern the "solar cylinder", of diameter 0.5 kpc (the size of our radial bin), centered on the Sun. 
%Our only  criterion thick disk...Thus, even a successful  comparison with

{\it Assuming } that the thick disk is the oldest ($>$9 Gyr) part of the disk, we found that the adopted radial migration scheme can reproduce quantitatively the main local properties of the thin and thick disk: metallicity-distributions, the characteristic "two-branch" behaviour of the local O/Fe vs Fe/H relation, local surface densities of stars (10 \mpc \ and 28 \mpc \ for the thick and thin disk, respectively). The thick disk extends up to $\sim$11 kpc and has a scale length of 1.8 kpc; this is consistent with recent evaluations (e.g. \citealt{Bovy2013} and references therein) and it is considerably shorter than the one of the thin disk, consistent with  the inside-out formation scheme. 

%We also show how, in this framework,  current and forthcoming spectroscopic observations can  constrain the nucleosynthesis yields of massive stars for the metallicity range of 0.1 \zs \ to 2-3 \zs.
Some of  the main results of the model relevant for this work appear in Fig. \ref{Fig:ModelGen1}.  The inside-out formation of the disk is evidenced by the evolution of both the infall rate profile and the stellar profile. The gaseous profile, mostly flat today, with a local surface density of $\sim$12 \mpc, is well reproduced. 
The profile of the molecular fraction $f_{Mol} = \frac{\rm H_2}{\rm (HI+H_2)}$ is well reproduced also, after the prescriptions of \cite{BlitzRos06}. As already mentioned,
this profile plays an important role in our model, because it determines the molecular profile and, thereoff, the profile of the star formation rate $\Psi (r) = f_{Mol}(r) \Sigma_{Gas}(r)$.  
%The latter (bottom left panel in Fig. \ref{Fig:ModelGen1}) is slightly flatter than the observations in the outer disk. being determined in that region by the slowly declining gas profile;  but in most of the disk, the SFR profile is well reproduced.

The profile of the SNIa/CCSN ratio  (bottom right panel) becomes steeper as one moves from the outer to the  inner Galaxy, because the SNIa rate - being a mixture of  old and young population objects - follows a combination of the stellar and gas profiles. The steep stellar profile increases substantially the SNIa/CCSN ratio in the inner disk; the outer disk, populated mostly by gas and young stars, has essentially SNIa belonging to the young stellar population, as the CCSN: as a  result, the SNIa/CCSN ratio in the outer disk is practically constant. Notice that similar results for the SNIa/CCSN ratio are obtained in the independent-ring model for the MW disk by 
\cite{Boissier2009}, both with the numerical prescription of \cite{Greggio05} and with an analytical prescription for the SNIa rate (their Fig. 11).  In the present study,  a fraction of SNIa - those belonging to the old stellar population -  is affected by radial migration, mostly in the region 3-12 kpc (see next paragraph).   All these features affect directly the resulting O and Fe profiles, as well as those of all other elements (see discussion in next section).

\begin{figure}
\begin{center}
\includegraphics[width=0.49\textwidth]{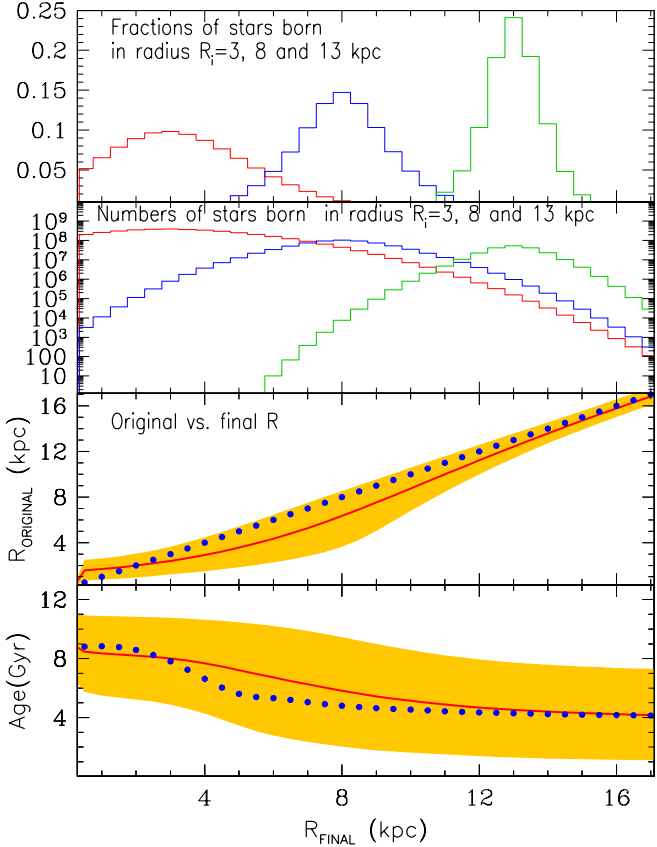}
\caption[]{{\it Top:} Fractions of stars born in radii $r_{Origin}$=3, 8 and 13 kpc and found in radius $r_{Final}$ at $t$=12 Gyr.
{\it Second row:} Numbers of stars born in annuli of radius $r_{Origin}$=3, 8 and 13 kpc and width $\Delta r$=0.5 kpc and found in radius $r_{Final}$ at $t$=12 Gyr.
{\it Third row:} Original vs final guiding radii of stars ({\it solid curve}); the shaded aerea includes $\pm$1 $\sigma$ values (i.e. from 16\% to 84\% of the stars) and the dotted diagonal line indicates the stellar guiding radii in the absence of radial migration. 
{\it Bottom:} Average age of stars vs Galactocentric radius  ({\it solid curve}); the shaded aerea includes $\pm$1 $\sigma$ values  and the {\it dotted} curve indicates the average age of stars formed {\it in-situ}.
}
\label{Fig:ModelGen2}
\end{center}
\end{figure}

Some aspects of the stellar radial migration of the model appear in Fig. \ref{Fig:ModelGen2}. The top panel displays the fraction of stars born in radius $r$ (in a radial bin of width $\Delta r=\pm$0.25 kpc) and found in the end in all radii. It can be seen that the action of the bar brings a large fraction of the stars of the inner disk in the outer regions: some stars born in $r$=3 kpc are found int he solar neighborhood in the end of the simulation. As we show in KPA14, these are the most metallic stars presently found in the solar neighborhood, with metallicities [Fe/H]$\sim$0.4 and they are 3-5 Gyr old. In contrast, a negligible fraction of the stars born in $r>$12 kpc reaches the solar vicinity\footnote{ For each "birth radius" there are approximately  as many stars migrating outwards as inwards. However, there are MORE STARS in the inner regions than in the outer ones, because of the exponentially declining outwards surface density profile. As a result, the outer regions receive more stars from the inner ones than what they send to them: the radial profile of the fraction of stars that each region receives from  the others is not symmetric but biased towards the inner regions, i.e. each disk annulus receives more stars from the inner regions than from the outer ones.}

The middle  panel of Fig. \ref{Fig:ModelGen2} displays the original vs final guiding radii of stars. Stars 
found in the radial range 3-14 kpc have been formed, on average,  inwards of their present position; the effect is most pronounced for stars in the region 6-9 kpc, where the average outwards displacement reaches $\Delta r\sim$1.5 kpc, and it is reduced to negligible values inside 4 kpc and outside 14 kpc. Also, the dispersion around those average values is large in the 6-9 kpc range and decreases outside it. Clearly, however, radial migration affects to some extent regions at all galactocentric distances. We stress that the extent of radial migration depends strongly on the adopted (time- and radius-dependent) diffusion coefficients and on the morphology of the disk galaxy: a larger amount of radial migration is expected in the case of barred disks, through the coupling of the bar to the spiral arms \citep{minchevfamaey2010}. 

 Finally, the bottom panel of Fig. \ref{Fig:ModelGen2} displays the average age of the stellar populations as function of Galactocentric radius, both for all the stars  and for the stars formed {\it in situ}. In the latter case, the average age varies little  with radius outside $r$=6 kpc, because the adopted profile of the infall timescale (Fig. \ref{Fig:ModelGen1}) varies little  in that region. However, radial migration brings older (on average) stars from the inner disk in intermediate radii: as a result, a clear age gradient is developed throughout the disk (solid curve in Fig.  \ref{Fig:ModelGen2} bottom).

As emphasized in KPA14, the impact of radial migration on the  properties of the disk is not intuitively straightforward, because migrating stars may return their gas (metal-rich, if originating from the inner disk or metal-poor if originating from the outer disk) in places far away from their home radius; this gas may affect the local metallicity and also fuel star formation (depending on the local star formation efficiency). The situation becomes even more complex in the case of a bar: the bar drives inwards gas - also fuelling  star formation in the inner disk  - which is more metal poor, in general, than the local gas and thus  the metallicity in the inner disk decreases; the overall result depends, however, also on the ratio of the local infall rate to the SFR and on the previous history of the disk (which determines the metallicity at a given time).
Oxygen is affected differently than Fe, because the main source of the latter, namely SNIa, is affected by radial migration, while the source of O (massive stars) is not.
 
\begin{figure}
\begin{center}
\includegraphics[width=0.49\textwidth]{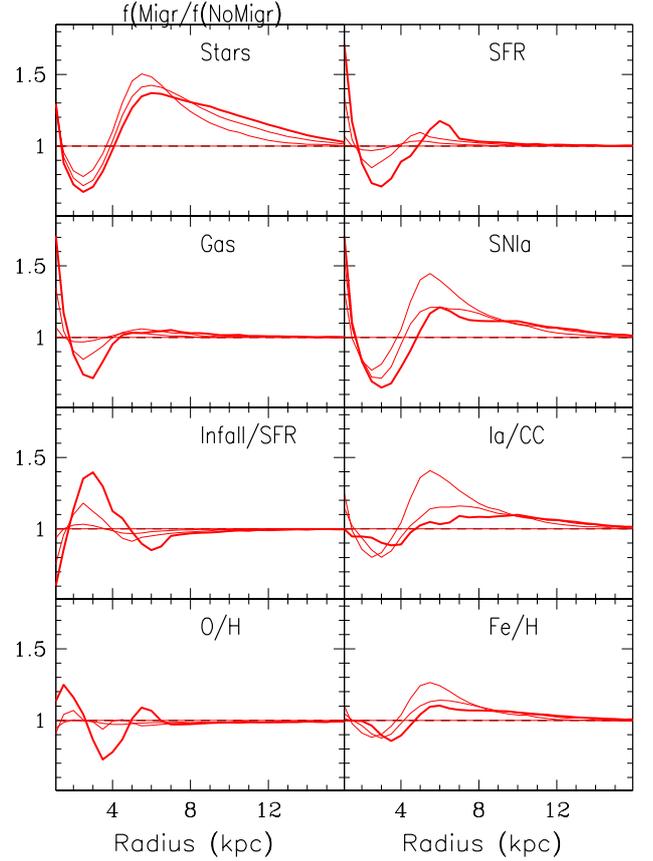}
\caption[]{{\it Top:} Comparison of various properties of our model, expressed as the ratio of a given quantity  in the case of radial migration to the same quantity when stars  do not migrate. From top to bottom, {\it left}: star surface density, gas surface density, gas fraction, O/H profile; {\it right}: SFR surface density, SNIa rate, ratio of SNIa/CCSN, Fe/H profile. In all panels, {\it thick curves} represent results at 12 Gyr and thin curves at 8 and 4 Gyr, respectively (as in Fig. \ref{Fig:ModelGen1}).
}
\label{Fig:ModelGen3}
\end{center}
\end{figure}

We attempt an illustration of  this complex behaviour 
in Fig. \ref{Fig:ModelGen3}, where we plot the ratio of several quantities of the model (with radial migration and radial inflow) to those same quantities obtained by an identical model (same boundary conditions, same SFR and infall rates) without radial migration or radial inflow.
One can easily see that the quantities affected mostly by radial migration
are the long-lived stars and SNIa: their radial profiles are affected over most of the disk.  The radial profiles of SNIa are affected to smaller extent than those of stars, because  a substantial fraction of SNIa results from a young population, unaffected by radial migration ($\sim$40\% of them explode within 1 Gyr after the formation of their progenitor system, see Fig. C1 in Appendix C of KPA2013); in contrast, most stars are low-mass and long-lived ($\sim$90\% by number for a normal IMF) and their population is affected by radial migration.

Gas is affected mainly not by radial migration, but by the radial inflow induced by the bar. Its surface density is depleted in the 2-4 kpc region and slightly increased outside it; notice that the latter increase is also due, to a small extent, to the gas returned to the ISM by the migrating, dying stars.
The evolution of the gas profile is reflected in the one of the SFR profile,
which is also affected by the radially dependent fraction $f_{H_2}$ of molecular gas: there is less SFR in the 2-4 kpc region than without
radial migration and gas inflow. That region is also the one corresponding to the peak of the infall rate during the late evolution of the disk (see Fig. \ref{Fig:ModelGen1}), and shows a higher infall/SFR ratio than the model with no radial migration. For those reasons (smaller SFR and higher dilution of the metallicity through the primordial infall), this region is found to have a lower O/H ratio than in the model with no radial migration. The decrease in the Fe/H ratio is smaller, because some Fe is contributed in those zones by SNIa migrating inwards. But the largest impact on the chemical evolution concerns the SNIa migrating outwards: they increase the  Fe content of the region between 5 and 8 kpc by $\sim$30\% after 4 Gyr and by $\sim$10\% in the end of the simulation.
As a result, the final Fe/H radial profile is somewhat steeper than in the model without radial migration. 

Overall, the effects of radial migration on the profiles of stars, SNIa, SNIa/CCSN ratio and Fe/H appear to become less important at late times. This
result appears counter-intuitive, at first sight, because more radial migration occurs at longer timescales (everything else kept equal). Our counter-intuitive result is due to the inside-out formation of the disk.
At early times, there are few stars  (and SNIa) in the outer Galaxyl: any radial transfer from the inner regions (where a large stellar population has been formed) to the outer ones, increases the surface density of the latter by a large amount. At late times, the stellar population   is in place all over the disk: the impact of radial migration from the inner to the outer disk (where a lot of stars are formed {\it in situ}) is proportionally smaller then.

KPA13 performed a similar exploration of the effects of radial migration for the case of a N-body+SPH simulation concerning a barred disk galaxy, evolving without gaseous infall (as a closed box). They found that, in that case,  the strong bar induced
a much larger amount of radial migration all over the disk, affecting particularly its outmost regions.
As discussed in KPA14, we adapted the description of the radial migration of that model to the one of the MW, by taking into account the size of the bar in the two cases. The smaller bar of the MW implies smaller extent of radial migration than in the barred disk of KPA13.

\section{Abundance evolution}
\label{sec:abundance-evol}

Our model includes the detailed evolution of 83 isotopes, from H to Zn. The abundances of the corresponding 32 elements are obtained at each time step by summing over the isotopic ones.

IN KPA14 we have found that the adopted parameters of the model (SFR efficiency, infall timescale, SNIa rate, IMF, etc.) allow us to reproduce quite well  the solar abundance of O and Fe for the {\it average 4.5 Gyr old star} in the solar neighborhood\footnote{We calculate the average metallicity $<[Z/H]>=\frac{\Sigma N_i[Z/H]_i}{\Sigma N_i}$ of stars of age $A$ found in zone $r$, where $N_i$ is the number of stars with metallicity $[Z/H]_i$ in that zone.} The average birth radius of those stars is found to be at a Galactocentric distance of $\sim$6.5 kpc, i.e.  $\sim$1.5 kpc inwards of the present day position of the Sun. This implies that the Sun is an average star of 4.5 Gyr in the Solar neighborhood as far as its chemistry is concerned, but such stars are not born {\it in situ}, they have migrated here from the inner disk. 

Regarding the other elements and isotopes, we find that the agreement with the solar abundances is quite good for most of them (to better than a factor of 2, see Fig.  17 in KPA14). In several cases, however, and in particular in the region of Sc to Mn, a clear underabundance is obtained with the adopted yields of  \cite{Nomoto2013}.  The reason of that disagreement is obviously a deficiency of the yields. A similar deficiency is obtained  when using those yields to study the evolution of the halo(see Fig. 10 in \cite{Nomoto2013}). To cure for that, we normalised the model results as to have the average abundances of the 4.5 Gyr old stars currently present in the solar neighborhood equal to their solar values, i.e. we corrected the stellar yields (integrated over the IMF) to various degrees (by 2\% for O, up to a factor of 4  for V).  In that way, we were able to compare the corresponding evolution of all the abundance ratios X/Fe vs Fe/H to observational data of two recent surveys \citep{Adibekyan2011,Bensby2014} concerning the local thin and thick disks separately.
We showed how such detailed comparisons in the future will provide valuable constraints to both stellar nucleosynthesis and chemical evolution models.

Here we extent  the investigation of the abundance evolution to the whole disk for all the elements of our model;
we leave the isotopic evolution for a future paper.
We use abundance data from various sources and different classes of objects: Cepheids, B-stars, HII regions, planetary nebulae (PN) and open clusters. The first three concern young objects: Cepheids have masses $>$3 \ms \ and, depending on their metallicity, they are younger than 300-400 Myr, while B-stars and HII regions are less than a few 10$^7$ Myr old;  on the other hand, PN and open clusters correspond to objects with ages up to several Gyr. The sources of the adopted data are presented in Table 1.

\begin{table}
\caption{\label{Tab:Data}{References for adopted data on Cepheids, B-stars, HII-regions, planetary nebulae (PN) and Open clusters (OC).}
}
\begin{tabular}{lccccc}
 \hline \hline
 Element  &  Cepheids &    B-stars & H-II & PN & OC\\
   \hline
C  & 1  & 4,5 &  &   & \\
N  & 1  & 4,5 & 6 & 7 & \\
O  & 1  & 4,5 & 6 & 7,8 &\\
Ne  &   &  &  & 7 & \\
Na &  1 &  &  &   & \\
Mg  & 1  & 4,5 &  & &  \\
Al  & 1  & 4,5 &  &  & \\
Si  & 1  & 4,5 &  &  & \\
S  &   & 4,5 & 6 & 7 & \\
Ar  &   &  &  & 7 & \\
Ca  & 1  &  &  &  & \\
Sc  & 1  &  &  &  & \\
Ti  & 1  &  &  &  & \\
V  &  1 &  &  &  & \\
Mn  &  1 &  &  &  & \\
Cr  & 1  &  &  &  & \\
Fe  & 1,2,3  &  &  & & 9,10,11 \\
Co  & 1  &  &  &  & \\
Ni  & 1  &  &  &  & \\
\hline
 
\end{tabular}  

1: \cite{Luck2011}, 2: \cite{Lemasle2013}, 3: \cite{Genovali2014},  4:\cite{Gummersbach1998}, 5: \cite{Daflon2004}, 6: \cite{Rudolph2006}, 7: \cite{Henry2004}, 8:\cite{Henry2010}, 9: \cite{Magrini2009}, 
10: \cite{Yong2012}, 11: \cite{Frinchaboy2013}
 \end{table}

\subsection{Oxygen profiles}
\label{sub:O-profiles}

Oxygen is a major product of the nucleosynthesis of massive stars (M$>$10 \ms). Being short-lived (lifetime $<$20 Myr), such stars have no time to migrate away from their birth sites (less than a 100 pc, as suggested by the fact that all CCSN localised up to now in external galaxies are within spiral arms and/or regions of active star formation). As a result, the radial O profile is not affected by radial migration. It is strongly affected, however, by gas radial inflows, as  found in numerous studies, with semi-analytical and N-body+SPH models, e.g. \cite{Mayor1981,Lacey1985,Tosi1988,Friedli1994,Portinari2000,Spitoni2011,Bilitewski2012,Cavichia2014}, etc. The results presented here depend directly on the adopted treatment of radial inflow, which corresponds to the action of the galactic bar, as described in Sec. \ref{sec:Model}. Among the aforementioned studies, only \cite{Portinari2000} and \cite{Cavichia2014} considered explicitly radial flows induced by the Galactic bar.

\begin{figure}
\begin{center}
\includegraphics[width=0.49\textwidth]{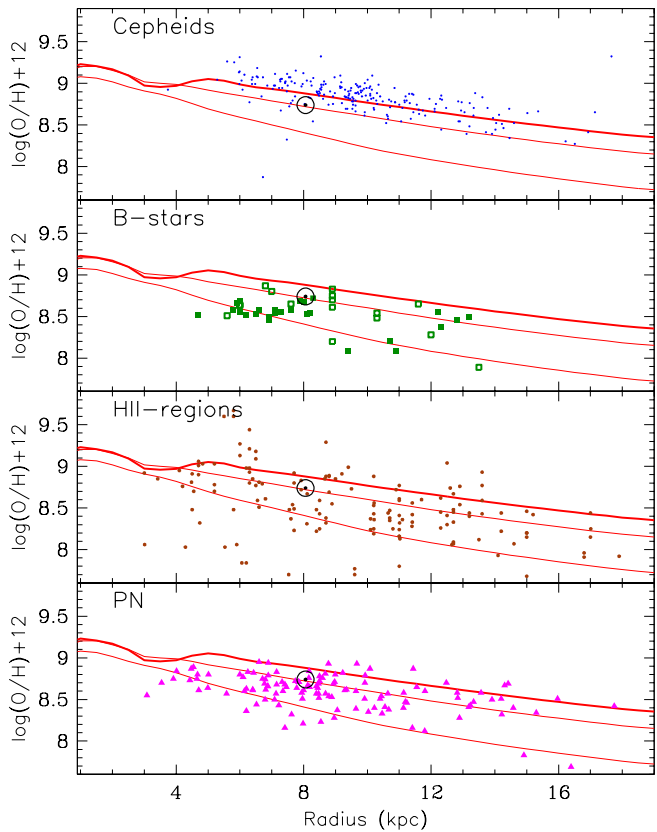}
\caption[]{Oxygen abundance profiles of Cepheids, B-stars, HII regions,  and planetary nebulae (PN); data sources are in Table 1. The adopted solar value
of log(O/H)+12=8.73 is from \cite{Asplund2009}. Model curves (from bottom to top, in all panels) represent the  gaseous abundance profile at time 4, 8 and 12 Gyr ({\it thick curve}), respectively.
}
\label{Fig:O-profiles}
\end{center}
\end{figure}
In Fig. \ref{Fig:O-profiles} we display the evolution of the gaseous profile of oxygen in our model at three different times (4, 8 and 12 Gyr, respectively). The O profile results from the joint action of three different factors:

- the inside-out formation of the disk, affected by both the adopted infall profile (shorter time-scale in the inner disk, Fig. \ref{Fig:ModelGen1}, top right) and the larger efficiency of star formation in the inner disk (because of the larger molecular fraction there, Fig. \ref{Fig:ModelGen1}, bottom left). 

- the radial inflow, which affects the gaseous profile and all the abundance profiles in the inner galaxy ($r<$6 kpc).

In the 2-4 kpc region,  the combination of  the action of the bar (which pushes gas partly towards the center and partly towards the outer disk) to the metal-poor infall leads to a local depression of O/H with respect to adjacent regions. This is due to the fact that the O rich gas pushed inwards and outwards from that region is replenished by the metal poor infall, the rate of which happens to be maximum in that region (Fig. \ref{Fig:ModelGen1}). In adjacent regions the effect is smaller, because the late infall rate is less intense there.

The disk beyond radius $r\sim$6 kpc is not affected by radial inflows. The resulting O/H profile is smoothly decreasing outwards, but it cannot be described by a single exponential over the whole radial range: its slope is steeper at small radii and flatter at larger ones. If it is fitted with a single exponential, then the slope depends on the radial range considered.
In this work, we shall consider as baseline values those in the range 5-14 kpc, where most of the observational data are available.

There are several shortcomings and uncertainties in the analysis of the observed O abundances of different types of objects across the Galactic disk, which are presented in the recent monograph of \cite{Stasinska2012}. The various surveys lead to widely different results for the O/H abundance gradient
$dlog(O/H)/dr$ (in \dxpc),  ranging from
small values (-0.023$\pm$0.06 in the PN sample of  \citet{Stanghellini2010} in the 2-17 kpc range) to high ones (-0.056$\pm$0.013 in the Cepheid sample
 of \cite{Luck2011} in the 5-16 kpc range).

In Fig.  \ref{Fig:O-profiles} we compare our results to the
data of some recent, representative surveys, of the aforementioned tracers.
We do not plot all the data on the same figure, since this would create confusion and increase artificially the scatter, because of systematic uncertainties between different analysis techniques (see \citet{Stasinska2012}. It can be seen that our results are in overall agreement with the various observations. 
There are practically no  data corresponding to the inner disk
($r<$4 kpc), where our model predicts lower values than in adjacent reasons, as discussed above. We notice, however, that in their study of PN in the direction of the Galactic bulge, \cite{Chiappini2009b} identified a subsample of 44 objects
which actually belong to the inner disk population  (a few kpc from the Galactic center) and have an average value of log(O/H)+12=8.52$\pm$0.23, i.e. less than expected from the extrapolation of the \cite{Henry2010} data for PN in that region. It is not yet clear whether that difference is due to systematic uncertainties between the two studies - \citet{Chiappini2009} collected
line intensities from the literature, unlike \citet{Henry2010} - or to a genuine decline of the oxygen profile in the inner disk, in qualitative agreement with our results.

\subsection{Iron profiles}
\label{sub:Fe-profiles}

\begin{figure}
\begin{center}
\includegraphics[width=0.49\textwidth]{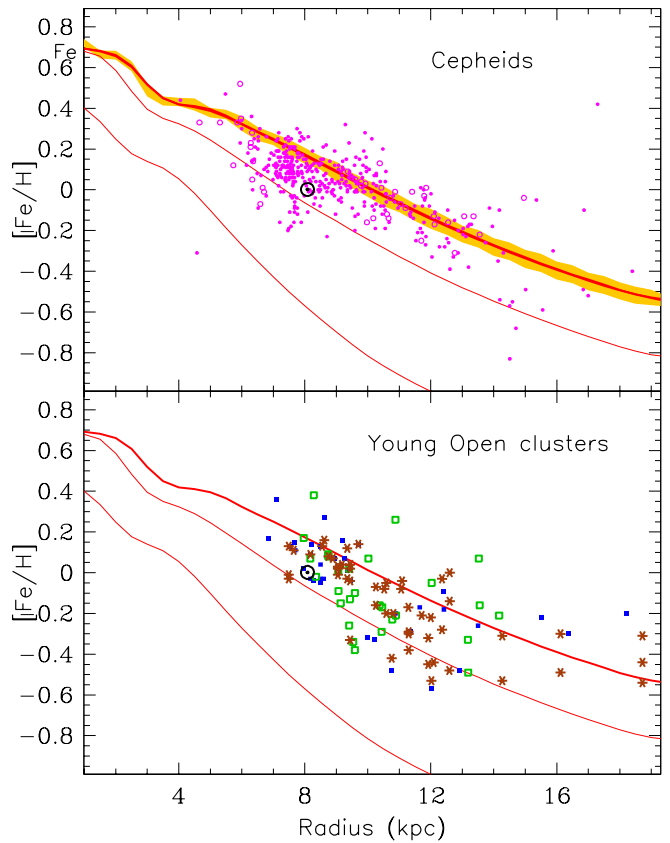}
\caption[]{Iron abundance profiles of  Cepheids (top) and open clusters (bottom). Data sources are provided in Table 1. Symbols in the upper panel correspond  to Ref. 3 and in the lower panel to Refs. 9 (blue filled squares), 10 (brown asterisks) and 11 (green open squares), respectively. Curves corespond to model results at time 4, 8 and 12 Gyr (thick curve), respectively). The thick red curve in the {\it top} panel corresponds to the average metallicity of a young stellar population of age 0.2$\pm$0.2 Gyr (Cepheids) and the shaded aerea represents the corresponding $\pm$1-$\sigma$ dispersion. 
}
\label{Fig:Fe-profiles}
\end{center}
\end{figure}

While O is exclusively  the product of massive stars, Fe  has two sources:
massive stars and thermonuclear supernovae.  
From the nucleosynthesis and chemical evolution points of view, Fe is then 
far more complicated to deal with than O. The reasons are

- The Fe yield of massive stars, exploding as CCSN, are difficult to calculate from first principles, since CCSN explosions are not well understood yet. 
Observations suggest that Fe yields depend on the energy of the explosion
(e.g. \cite{Hamuy2003})  but this parameter is not systematically taken into account in yield calculations.

- Most of solar Fe appears to come not from CCSN but from SNIa (on the basis of the observed decline of O/Fe in disk stars), but is is difficult to relate in a unique way the  rate of SNIa to that of CCSN (see, however, Appendix C in KPA2014).

Radial migration introduces one more layer of complexity in the story of Fe.
As shown in \cite{KPA2013}, a fraction of SNIa - mainly those resulting
from the oldest stellar populations - may be affected by radial migration as single stars are. The effect is quite important in the disk of the simulation of \cite{KPA2013}, which displays a long and strong bar,
its semi-major axis reaching in the last evolutionary stages between 6 and  8 kpc. 

In this work, the effect of radial migration appears to be rather small for SNIa and Fe production (see right panels in Fig. \ref{Fig:ModelGen3}). The reason
is that star formation proceeds at a quasi-constant rate over most of the disk, creating a large number of SNIa {\it at late times}; in those conditions, the migration of some old SNIa progenitors  from the inner disk, modifies little the situation (and, in any case not beyond a galactocentric radius of $r\sim$12 kpc). In contrast, in the simulation of \cite{KPA2013}, there is very little star formation in the whole disk after the first couple of Gyr, due to the lack of accreting gas; as a result, the radial migration of SNIa progenitors from the inner disk during the subsequent 8 Gyr of evolution (under the action of the strong bar), increases considerably the
SNIa population and the concomitant Fe production in the outer disk.

Our results for the Fe/H profile are displayed in Fig. \ref{Fig:Fe-profiles}, at three different times: 4, 8 ad 12 Gyr, respectively. In the latter case, we display in the upper panel the average metallicity of  stars aged between 0 and 0.4 Gyr, (i.e. covering the range of Cepheid ages), along with the corresponding range of $\pm$1-$\sigma$ values. Being young objects, Cepheids have no time to migrate away from their birth places and theirs radial profile after migration (displayed here) is practically the same as the one they have at their birth.
The upper panel shows clearly that radial migration introduces very little dispersion in [Fe/H]  for such young objects. If the observed dispersion in the sample of \cite{Genovali2014} is real and not due to  measurement errorss, then its origin should be due to other factors (e.g. uncertainties in radial distance estimates, azimuthal variation of Fe/H, etc.)

A few  other features of the Fe/H profile in Fig. \ref{Fig:Fe-profiles} are worth noticing:

- The profile flattens off in the 3-5 kpc region, instead of presenting there a decrease, as the O profile does. The reason is that in this region, at late times there is a population of old progenitors of SNIa (formed early on) which produces some Fe and compensates for the deficiency  of CCSN there; this is not the case for O, as we discussed in the previous section. This contribution of old SNIa turns out to be sufficient to smooth the Fe/H profile in that region.

- Outside that region, the Fe/H profile decreases rather steeply, more steeply in any case than the corresponding O profile. The reason is that the ratio of SNIa/CCSN is always higher in the inner disk than in the outer disk (see right bottom panel of Fig. \ref{Fig:ModelGen1}), because the former has both an old and a young population of progenitors, while the latter has only a young population. As a result, Fe production is more important proportionally to the one of O in the inner disk, and the resulting Fe profile is steeper.

- In the outer disk, the Fe profile is less steeply decreasing, for the same reasons as the O profile (see previous section), namely the star formation efficiency of the adopted prescription for the SFR. Again, the profile cannot be described by a single exponential.

Comparison to observations is reasonably good, given the dispersion in the data.
In the case of the open clusters, dispersion appears to be even larger than in the case of Cepheids; here, however, the (poorly determined) age of the clusters, which covers a range of several Gyr, certainly contributes to this effect.
Although our Fe profile flattens in the outer disk, we never obtain a quasi-constant Fe/H abundance beyond $r\sim$15 kpc, in contrast to the observational findings  of \cite{Yong2012} and \cite{Heiter2014} for open clusters.

Compared to other models in the literature, our results are closer to those of \cite{Naab2006}, as far as the overall Fe profile is concerned, which is also
steeper in the inner disk and progressively flattens outwards. Manifestly, this is due to the similar dependence of the SFR on radius in the two cases, steeper in the inner disk and flatter in the outer disk. In our case, this dependence results from the adopted SFR proportional to the molecular gas (Fig. 1, left panels), while in the case of \cite{Naab2006} it results from the adopted law $SFR \propto\Sigma_{GAS}/\tau_{DYN}$, with the dynamical timescale $\tau_{DYN}\propto r$ for a flat rotation curve: the factor $1/\tau_{DYN}$ varies considerably in the inner disk and much less outside 10 kpc.
On the other hand, the model of \cite{Magrini2009} does produce a nearly flat Fe profile in the outer disk -
even for intermediate ages -  presumably through  some appropriate combination of SFR and infall rate there

\subsection{Evolution of abundance profiles}
\label{subsec:Evolution_profiles} 
 
The abundance profiles of gas and stars depend on the  interplay between star formation, infall, radial inflow and radial migration of stars. 
Observations of the final profiles alone can hardly shed light on this complex interplay. The history of the abundance profiles, if observed through some tracer of well determined age, could help in that respect: indeed, some semi-analytical models predict 
gradients steeper in the past (e.g. \citet{Hou2000}) while others predict that gradients are flatter for older objects (e.g. \cite{Chiappini01}). As discussed in \cite{Pilkington2012}, who surveyed 25 models (both semi-analytical and with N-body+SPH codes), models may also differ widely as to the rate of change of the gradients with time (see \cite{Gibson2013} for an update).  

Observations of planetary nebulae of different age classes suggested that O 
gradients were steeper in the past \citep{Maciel2009}. However, the systematic uncertainties affecting age and distance estimates of those objects make it difficult to use them as tracers of the past gradient evolution
at present.  Even worse, radial migration modifies considerably the radial profiles of stellar populations, as found in \cite{Roskar2008}, by mixing metal-rich stars from the inner regions in the outer disk. In those conditions, it becomes difficult to use abundance profiles of old objects to infer directly the 
chemical evolution history of a galactic disk. Still, such observations, combined to other data (e.g. photometry profiles, stellar gradients as a function of distance from the galactic plane, etc.) and to appropriate models - taking properly into account the observational biases -  may provide valuable information on the history of the Galaxy.

\begin{figure}
\begin{center}
\includegraphics[width=0.49\textwidth]{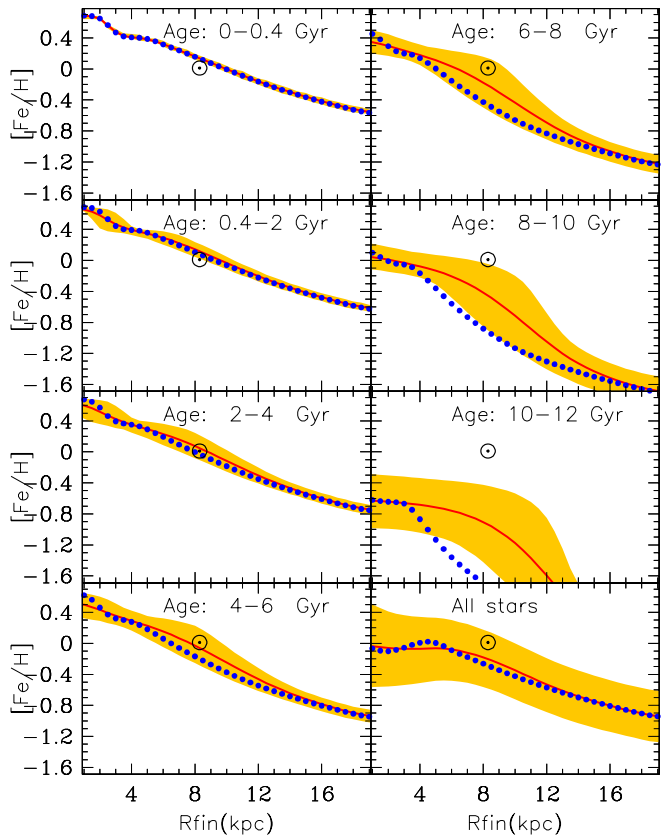}
\caption[]{{\it Top} Evolution of the  Fe profile for stars of different ages, born {\it in situ} ({\it dotted} curves) and presently found at radius $R_{fin}$ ({\it solid}); the {\it shaded} aerea indicates the $\pm$1 $\sigma$ range of values. 
}
\label{Fig:FeMigvsNoMig}
\end{center}
\end{figure}

In Fig. \ref{Fig:FeMigvsNoMig} we display the evolution of the Fe profiles of our model for all the stars ever born (right bottom panel) and for stars of different age ranges. We show the average metallicity for stars formed {\it in situ} (dotted curves)\footnote{This is not the same thing as the average gas metallicity during the corresponding time interval: the average stellar metallicity is weighted with the star formation rate during that period, whereas the average gas metallicity is not.} and for all stars found in radius $r$ at the end of the simulation (solid curves); the latter population has been affected by radial migration. It can be seen that for the younger stars (up to 4 Gyr old) the differences between  the corresponding profiles is small, for two reasons: i) radial migration does not have time to shuffle stars away from their birth places, and (most importantly) ii) at late times, the abundance profile is flatter than in earlier period, so that even an efficient radial migration cannot produce a large effect, because the abundance differences between different radii are small in any case. Still, radial migration increases steadily the dispersion in metallicity with age at all radii.

For stars older than the Sun, the effect of radial migration on the abundance profiles becomes more and more important, as it makes the profiles appear today flatter than they were at the time of the stellar birth, and flatter than the ones of younger stellar populations (despite the fact that the corresponding gaseous profile was steeper in the past). In particular, the oldest stars (presumably belonging to the thick disk) have a quasi-flat profile in the inner region, extending up to 6 kpc; beyond 9-10 kpc, however, the corresponding metallicity drops rapidly to values characteristic of halo stars.

The impact of radial migration on the past abundance profiles of a galactic disk was first identified by \cite{Roskar2008}: in their Fig. 2 they show how the older stars of their simulation ($>$5 Gyr) have a quasi-flat metallicity profile throughout the disk. 
Although it is difficult to compare directly our results with other models of similar scope, because of the many different assumptions involved (see KPA2014 for a brief description of the differences between the models of \cite{SB2009}, \cite{Minchev2013} and KPA2014), we attempt here such a comparison to the results
of \cite{Minchev2014}. In their Fig. 9 (top left), they provide results for the stellar abundances of practically all stars of their model (found in the end of the simulation within a distance of 3 kpc from the galactic plane), as function of galactocentric radius and stellar age. There are differences and  similarities with our results, but it is not clear whether the latter are due to similarities in the models or to different boundary conditions. In particular,
 they also find that the older stars have a flatter abundance profile
than the younger ones; however, this is probably due not to radial migration, but to the fact that in their case the abundance profile of gas is also flatter in early time than lately, a characteristic feature of the model of \cite{Chiappini01}. In their case, dispersion in metallicity is  more important for older stars than for young ones, as in our case and probably for the same reason, i.e. more time for radial migration being available to older objects and/or larger epicyclic motions; however, this dispersion appears to extend further outwards for the younger stellar population, whereas the opposite is obtained in our case.
Finally, the older stars in their simulation display a quasi-flat abundance profile all over the disk, whereas our corresponding metallicity profiles plummet beyond 9-10 kpc. This difference is simply due to the fact that we start our simulation with gas of primordial composition, whereas they adopt an initial metallicity of 0.1 \zs. Such differences may have negligible impact in some cases (i.e. for almost any observable concerning the solar neighborhood), but turn out to be crucial in others.

\begin{figure}
\begin{center}
\includegraphics[width=0.49\textwidth]{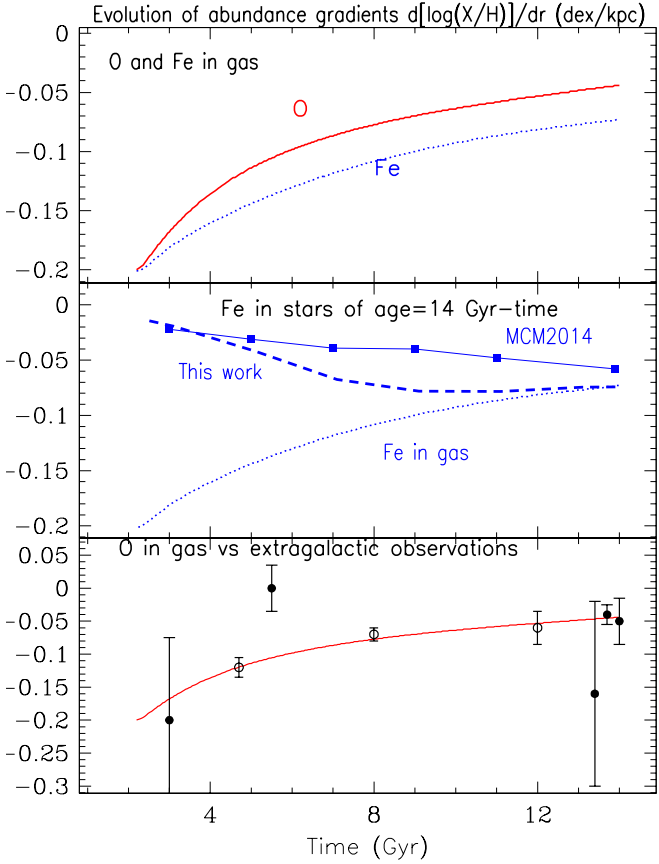}
\caption[]{{\it Top}: Evolution of the gaseous O and Fe abundance gradients. {\it Middle}: The evolution of the gaseous Fe gradient ({\it dotted, same as in top panel})  is compared to the evolution of the stellar Fe gradient,  as it appears today after stellar migration ({\it thick dashed}). The squares, connected by solid segments represent the results of \cite{Minchev2014} (their Table 2) for all stars found within distance $Z<$3 kpc from the galactic plane. {\it Bottom}: The evolution of the O abundance gradient in the gas (same as in top panel) is compared to the extragalactic data compiled from \citet{Jones2013}; notice that the three {\it open symbols} correspond to an evaluation of MW data, which is  superseded by the recent one of  \cite{Maciel2013}, who find no clear indications for an evolution of the abundance gradient. .
}
\label{Fig:Evol_grad}
\end{center}
\end{figure}

We show the evolution of the O and Fe abundance gradients in 
Fig. \ref{Fig:Evol_grad}. In the top panel we display the evolution of the
gradients in the gaseous phase. As already discussed, gaseous gradients decrease in absolute value with time, i.e. the abundance profiles become flatter with time (at least in the framework of this type of models). \cite{Hou2000} performed the first comparison between the evolution of the O and Fe abundance gradients and found that Fe gradients are steeper than those of O - by $\sim$0.1 dex - because of the role of SNIa: the ratio of SNIa/CCSN is larger in the inner disk than in the outer one (see Fig. \ref{Fig:ModelGen1}). We confirm this result here, 
but we obtain a larger difference between the two gradients - $\sim$0.25 dex -  because of the more efficient star formation in the inner disk and the role of radial migration: the latter  increases by $\sim$10\% the abundance of Fe, but not the one of O,  in the region outside 6 kpc (see bottom panels in Fig.  \ref{Fig:ModelGen3}).

As discussed in the previous paragraphs, radial migration modifies the {\it presently observed} evolution of stellar profiles. This is illustrated in the middle panel of Fig. \ref{Fig:Evol_grad}, where the evolution of the Fe gradient
in the gas is compared to the Fe gradient of stellar population as a function of their age. The gradients are the same for the last $\sim$2 Gyr
(see also Fig. \ref{Fig:FeMigvsNoMig}) but beyond that age the two curves start deviating: the one corresponding to the stellar population becomes flatter with age. For the oldest stars, the gradient is close to zero, i.e. the abundance profile is practically flat. Our results are qualitatively similar to those of \cite{Minchev2014}, also displayed in Fig. \ref{Fig:Evol_grad}, although the evolution is milder in the their case.

It is difficult to compare directly the "age effect" of radial migration on the abundance profile to observations, because of the uncertainties in stellar age estimates. However, there
is an indirect way, through the fact that older stars are, on average located further away from the plane of the disk than younger ones, because of the increase in the vertical velocity dispersion with stellar age. Thus,
analysing a sample of old, main sequence stars belonging to the thin and thick disks from the SEGUE survey, \cite{Cheng2012a} find that the Fe gradient in the 
region 6$< r$(kpc)$<$16 increase  from -0.065 \dxpc at 
vertical distance from the plane $Z$=0.2 kpc to a positive value at $Z>$1 kpc;
this is a clear signature of older stellar populations having flatter abundance profiles, as found in \cite{Roskar2008,Minchev2014} and in this work.
However, our model lacks the vertical dimension to the galactic plane and thus we cannot compare directly to the data of \cite{Cheng2012a}: at every distance from the plane there is a mixture of stellar populations, the contributions of older stars increasing with the distance.  Our results presented in Fig. \ref{Fig:Evol_grad} (middle panel) include all stars found today between galactocentric radii of 4 and 11 kpc.  Qualitatively, they are in agreement with the observations, since they suggest a gradient close to nul for the oldest stars and close to -0.07 dex/kpc for the youngest ones. A detailed comparison to the observations would require a model including the $z$ dimension (vertical to the plane), including the observational biases, i.e. slices at appropriate distances from the plane, as in \cite{Minchev2014}.
Alternatively, such a comparison would be possible if a volume limited sample with accurate stellar ages were available.

The bottom panel of Fig. \ref{Fig:Evol_grad} illustrates another way of comparing model results to observations of abundance gradient evolution. The results concern the evolution of the oxygen abundance gradient in the gas (as in top panel). The data are from observations of oxygen in high redshift lensed disk galaxies, from the recent compilation of \cite{Jones2013}; they find that the metallicity gradients flatten with time, by a factor of 2.6$\pm$0.9, on average, between redshifts 2.2 and 0, although they acknowledge that the discrepancy with the MASSIV data - the highest data point in the bottom panel of Fig. \ref{Fig:Evol_grad} - warrants further investigation. Barring that puzzling discrepancy, we find a rather fair agreement of the high redshift data with our results. It should be stressed, however, that the comparison may not be meaningful after all, because its not clear whether those isolated high redshift systems are progenitors of MW-like disks. 
%Still, those measurements provide strong support to models predicting that gaseous abundance profiles of galactic disks flatten with time, as this one. 

\begin{figure}
\begin{center}
\includegraphics[width=0.49\textwidth]{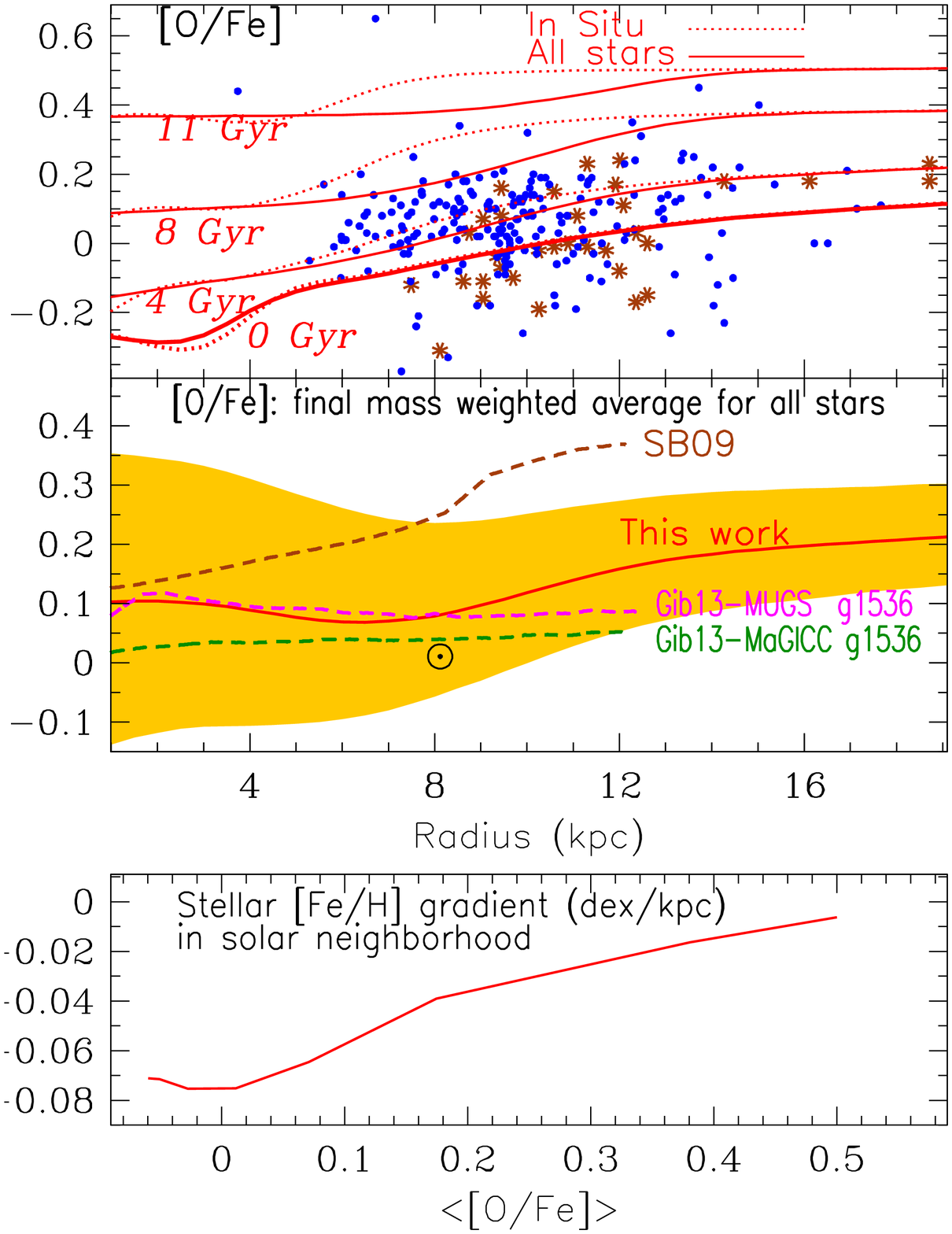}
\caption[]{{\it Top} Evolution of the O/Fe profile. Data are  for Cepheids ({\it filed circles}, from\cite{Luck2011}) and for open clusters (\cite{Yong2012}, {\it asterisks} and \cite{Frinchaboy2013}, {\it squares}). {\it Dotted}  curves correspond to model stars of average ages 11, 8, 4 and 0.2 Gyr (from top to bottom)  formed {\it in situ}  and {\it solid} curves to stellar populations of the same age  found today at radius $r$.
{\it Middle:} Mass weighted stellar [O/Fe] profile vs radius. The {\it solid}
red curve indicates our results and the {\it shaded aerea}  the corrsponding $\pm$1-$\sigma$ range. The {\it dashed} curves are from Fig. 2 of \cite{Gibson2013} and indicate results of \cite{SB2009} and \cite{Gibson2013}, the latter obtained with two different models (see text).
{\it Bottom}: The stellar Fe/H gradient of stars found today in the region 5-11 kpc is plotted vs. the average [O/Fe]
ratio of those stars (see text). 
}
\label{Fig:F_OvsFe_evol}
\end{center}
\end{figure}

\subsection{O/Fe profile}
\label{sub:OFe}
The variation of  the O/Fe ratio  provides important information on the
evolutionary status of a galaxian system: high O/Fe values (typically $\sim$3 times solar) indicate a  chemically young system, enriched only by the ejecta
of CCSN, while $\sim$solar values indicate systems several Gyr old, enriched also by SNIa. The transition from high O/Fe (and, more generally, high $\alpha$/Fe)
to low O/Fe values constitutes one of the key tracers of the chemical evolution
of the local Galaxy (the halo to disk transition) and of nearby dwarf galaxies as well.

In the case of the MW disk,the O/Fe ratio is expected to vary, from high values in the "young" outer disk, to lower ones in the older inner disk, in the framework of the inside-out formation scheme. In Fig. \ref{Fig:F_OvsFe_evol} (top panel), we plot the O/Fe radial profile for stellar populations of ages 11, 8, 4 and 0 Gyr (from top to bottom), for all the stars found in a given region in the end of the simulation (solid curves) and for stars  formed {in situ} (dotted curves). The decrease (with time) of O/Fe occurs first in the inner galaxy and 
 progressively moves outwards. The youngest objects have [O/Fe]$\sim$0.1 in the outer disk and $\sim$-0.25 in the innermost regions, whereas for the oldest objects the ratio  varies from 0.5 to 0.4. As in the case of the Fe/H profile, radial migration  modifies the O/Fe profiles by bringing evolved stellar populations (of lower O/Fe) into outer regions; the effect is more important for the oldest stars and affects the region between 4 and 12 kpc.
 
Similar results, at least qualitatively, appear in Fig. 9 (bottom left panel) of \cite{Minchev2014}, where the [Mg/Fe] profile is plotted for stars of different ages. Mg being an $\alpha$ element, a comparison to the O/Fe profile is meaningful \footnote{Notice, however, that halo stars appear to have, in general  higher [O/Fe] than [Mg/Fe] ratios at a given metallicity.}. They obtain a variation of [Mg/Fe] for their youngest stars ranging from $\sim$-0.16 at $r$=6 kpc to $\sim$0.1 at $r$=16 kpc, as well as a flat Mg/Fe profile for the oldest stars; both results are in fair agreement with ours.

Fig. \ref {Fig:F_OvsFe_evol} (top panel) displays observational data from Cepheids and open clusters of various ages. As with the corresponding Fe/H profile of Fig. \ref{Fig:Fe-profiles}, the dispersion in the O/Fe ratio  at every galactocentric radius is quite large and cannot be explained with our models; radial migration can play only a marginal role in that respect, for so young objects. No clear trend with radius appear in the case of Cepheids,  while the data  for open clusters are marginally consistent with such a trend, as discussed in \cite{Yong2012}.

In the middle panel of Fig. \ref{Fig:F_OvsFe_evol}, we plot the results for the mass weighted average of all stars at the end of our simulation.
We obtain a rather flat profile in the inner disk. The average [O/Fe]$\sim$0.1 in that region corresponds to stars older than 8 Gyr, as inferred through a comparison to the upper panel, which have been mixed throughout the inner disk by radial migration. In the outer disk, less affected by radial migration, the average [O/Fe] ratio  increases  slowly but steadily, up to a value of 0.2. The overall 1-$\sigma$ dispersion is much larger in the inner disk than in the outer one, as indicated by the shaded aerea. We compare our results to the corresponding ones reported in
Fig. 2 of \cite{Gibson2013} and reproduced in the middle panel of our Fig. \ref{Fig:F_OvsFe_evol}. The semi-analytical model of \cite{SB2009} displays
a much steeper slope of [O/Fe] vs. radius, which may result from less radial mixing than in our case or from a much larger gradient of stars formed {\it in situ}. We think that both reasons contribute to the difference with our results,  taking into account that \cite{SB2009} obtain quite large Fe gradients (dlog(Fe/H)/dr$\sim$-0.1 \dxpc) and that they consider radial mixing induced only by the transient spiral mechanism of \cite{SellwoodBinney2002} and not by the more efficient bar-spiral interaction of our model. On the other hand,  both models of \cite{Gibson2013} display a very flat profile of [O/Fe] over the whole disk,
which is rather difficult to understand, in view of the enhanced SNIa/CCSN ratio in the inner disk expected from inside-out formation schemes (as discussed in Sec. 2), unless if a very efficient radial mixing occurs for the stars over the whole disk. It is clear, however, that for different reasons, not necessarily well analysed yet, different models make different predictions for the profiles of metallicity and of various abundance ratios and only observations will help clarifying the situation.

\begin{figure}
\begin{center}
\includegraphics[width=0.49\textwidth]{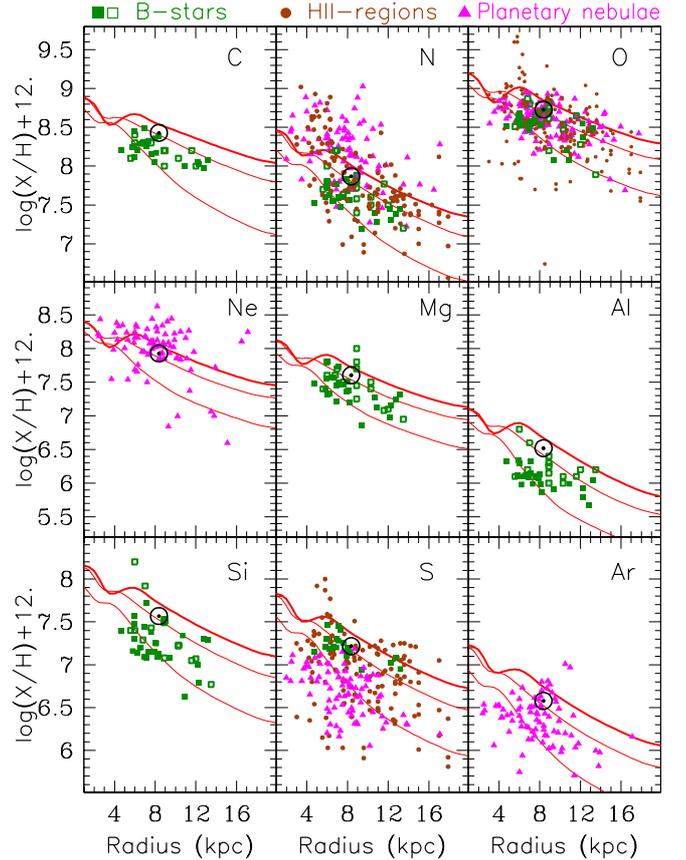}
\caption[]{Abundance profiles from C to Ar and comparison to data from PN (magenta {\it triangles}), B-stars (green {\it squares}) and HII-regions (brown {\it dots}); data sources are provided in Table 1. Model curves correspond to gaseous profiles at time 4, 8 and 12 Gyr ({\it thick}), respectively.
}
\label{Fig:F_Other_other}
\end{center}
\end{figure}

Finally, the bottom panel of Fig. \ref{Fig:F_OvsFe_evol} illustrates another use of the O/Fe ratio to probe the evolution of the Galactic disk. As found in KPA2014, the O/Fe ratio declines monotonically with time and displays very little dispersion from radial mixing at any age. It constitutes thus a natural "chronometer" as argued in \cite{Bovy2012} and it can be used in cases where stellar ages are not known or accurately measured.
\cite{Toyouchi2014} have analysed 18500 disk stars from the SDSS and HARPS surveys and plotted the Fe/H gradients as a function of the [$\alpha$/Fe]
values of the corresponding stellar populations. They found that, starting
with youngest stars (lowest [$\alpha$/Fe] values), the gradient first decreases, i.e. it becomes more negative and then increases, reaching positive values. However, they found large systematic differences between the samples of the two surveys (concerning the absolute values of the gradients and the turning points in [$\alpha$/Fe]). We display our results in the 
bottom panel of Fig. \ref{Fig:F_OvsFe_evol}, showing a qualitative agreement with the findings of \cite{Toyouchi2014}: starting with the youngest stars, the Fe/H gradient shows first a small decline, as slightly older stellar populations are probed (with a steeper Fe gradient because too young to be affected by radial migration); then, older populations are probed, more and more affected by radial migration and displaying flatter Fe/H profiles (as also indicated in the middle panel of our Fig. \ref{Fig:Evol_grad}). The oldest stars have a nearly flat Fe/H profile, but we never find a positive gradient, as \cite{Toyouchi2014} do. We stress again that a meaningful comparison to observations should involve models properly accounting for observational biases.

%\begin{figure}
%\begin{center}
%\includegraphics[width=0.49\textwidth]{F_Model_OFeMigsVsInSitu.eps}
%\caption[]{{\it Top} Evolution of the O/Fe profile. Data are from \cite{Luck2011} for Cepheids (small dots) and
% from ...... (big dots). {\it Dotted} magenta curves correspond to gas profiles at time 4, 8 and 12 Gyr ({\it thick}) %after the beginning,  respectively. {\it  Solid} red curves indicate the average O/Fe profile of stars of ages %0.2$\pm$0.2 Gyr ({\it thick}, 4$\pm$1 Gyr and 8$\pm$1 Gyr, respectively. CHECK THAT THE STELLAR AGES %CORRESPOND TO GASEOUS ONES
%}
%\label{Fig:F_OvsFe_evol}
%\end{center}
%\end{figure}

\subsection{Other elements}
\label{sub:Other}

Oxygen and iron are the most frequently observed elements in the solar neighborhood and the MW disk. Their abundances (as a function of time and/or space) constitute important constraints on the chemical evolution of the Galaxy. The other elements play a marginal role in that respect: their observations mainly serve to support conclusions obtained through the observations of O and Fe or to constrain stellar nucleosynthesis models.

In Fig. \ref{Fig:F_Other_other} we present our results for eight more elements, with abundance profiles derived from observations of B-stars, HII-regions and planetary nebulae.
In most cases, available data are for the region 4-12 kpc, with the exceptions of N and S (and, of course, O) where observations of H-II regions and PN extend up to 17 kpc. The inclusion of PN,  presumably covering a wide range of ages, increases considerably the dispersion at every radius.

\begin{figure*}
\begin{center}
\includegraphics[angle=-90,width=0.99\textwidth]{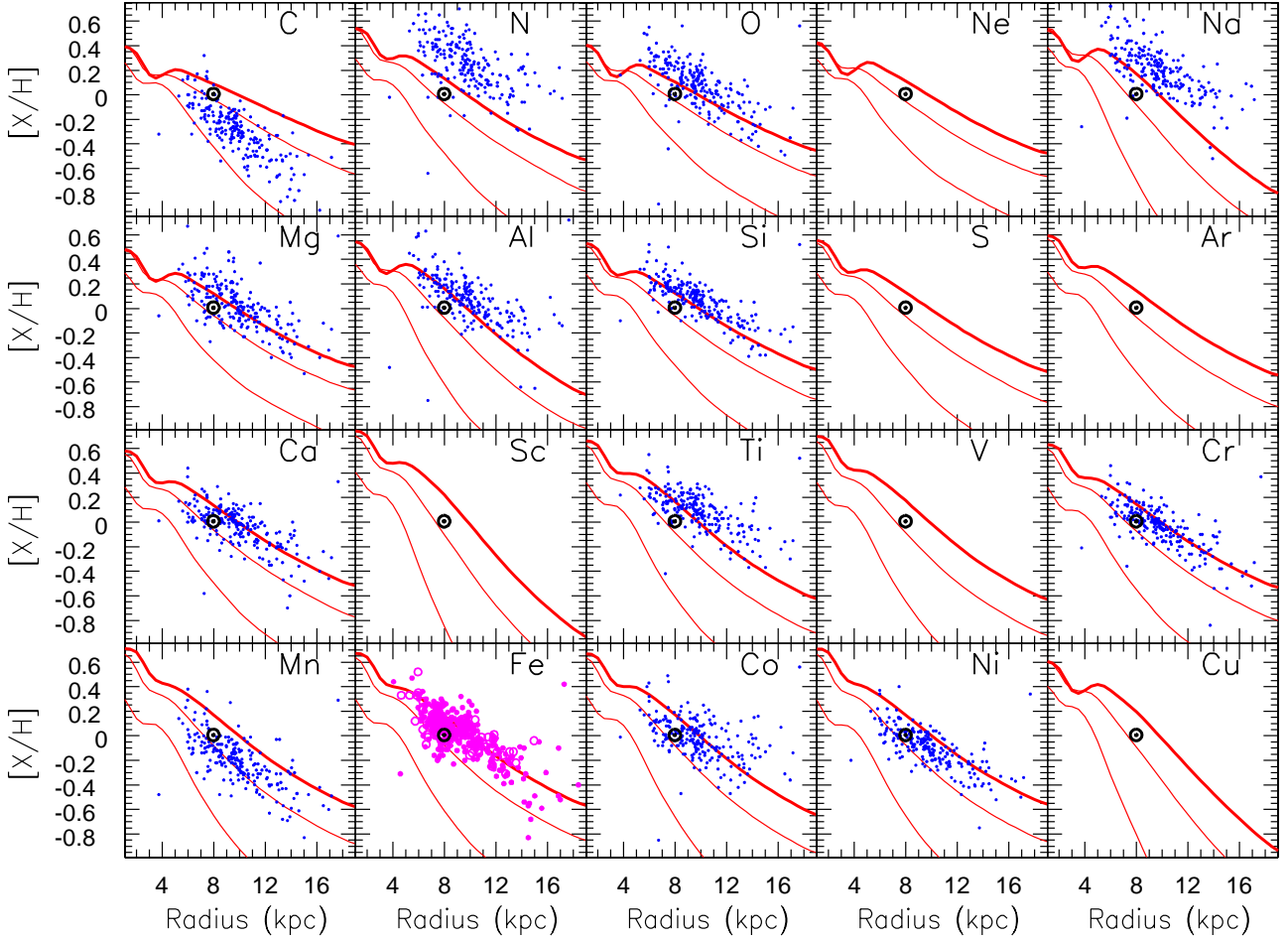}
\caption[]{Abundance profiles from C to Zn and comparison to Cepheid data (from \cite{Luck2011}, except for Fe data, provided in \cite{Genovali2014}). Model curves correspond to the gaseous profiles after 4, 8 and 12 ({\it thick curve}) Gyr, respectively.
}
\label{Fig:F_Other_Cepheids}
\end{center}
\end{figure*}

Among the 9 elements of Fig. \ref{Fig:F_Other_other}, those heavier than N are almost exclusively products of massive stars; Si and heavier elements receive  a small, but non negligible contribution from SNIa, that we take into account through the adopted yields of SNIa from \cite{Iwamoto99}. All those elements are produced as "primaries", i.e. their stellar yields depend little on the initial metallicity of the stars.  C and N have a complex nucleosynthetic origin, since they are produced both by massive and and intermediate mass stars.
Their yields from massive stars are sensitive to yet poorly  understood (and metallicity-dependent) stellar properties, as mass loss and rotation. In lower mass stars, C is produced in the shell He-burning of the AGB phase and ejected in the ISM through the 3d dredge-up, while N may be produced in the bottom of the convective envelope of the AGB star ("Hot-bottom burning") at the expense of C. N is, in principle, a "secondary" element (being synthesized from the initial C and O during the CNO cycle, its yield depends on the initial metallicity); yet, it may be produced essentially as primary in the "hot-bottom burning" of intermediate mass AGB stars  and in fast rotating massive stars.
 
The yields from \cite{Nomoto2013} that we adopted in this work, include yields from low-mass stars from \cite{Karakas2010}, but the ones for massive star concern  non-rotatings stars. Given the complexity of the nucleosynthesis of C and N, we consider that a dedicated study for the evolution of those two elements would be necessary and we do not attempt it here.

The results displayed in Fig. \ref{Fig:F_Other_other} present similar features to those already discussed for O, namely a flattening of the abundance profiles with time (due to the inside-out formation) and a slightly hollow profile in the bar region at late times (due to the combined action of the bar and of metal-poor infall).
Available data, however, display either too large differences (e.g. between B-stars and PN for S) or too large dispersion (in the case of PN data) or they are too scarce (e.g. for C) to allow for any serious constrains, either on the model or on the yields.  
(Fig. \ref{Fig:Evol_grad}, middle panel)

In Fig. \ref{Fig:F_Other_Cepheids} we compare our results for all elements between C and Ni to a  homogeneous data set for Cepheids \citep{Luck2011}, large enough for a statistically meaningful comparison with models. Still, Cepheids are relatively massive ($>$3 \ms) and evolved stars, having gone through the first dredge-up. This implies that they are expected to exhibit large amounts of N at their surface, formed {\it in situ} at the expense
of C, and perhaps of O; Na is also possibly affected by H-burning in those stars. In consequence, none of those four  elements observed in Cepheids can be used as tracer of the chemical evolution of the Galaxy (but they may certainly be used as probes of the internal nucleosynthesis of Cepheids).

\begin{figure*}
\begin{center}
\includegraphics[angle=-90, width=\textwidth, bb=243 20 593 780, clip]{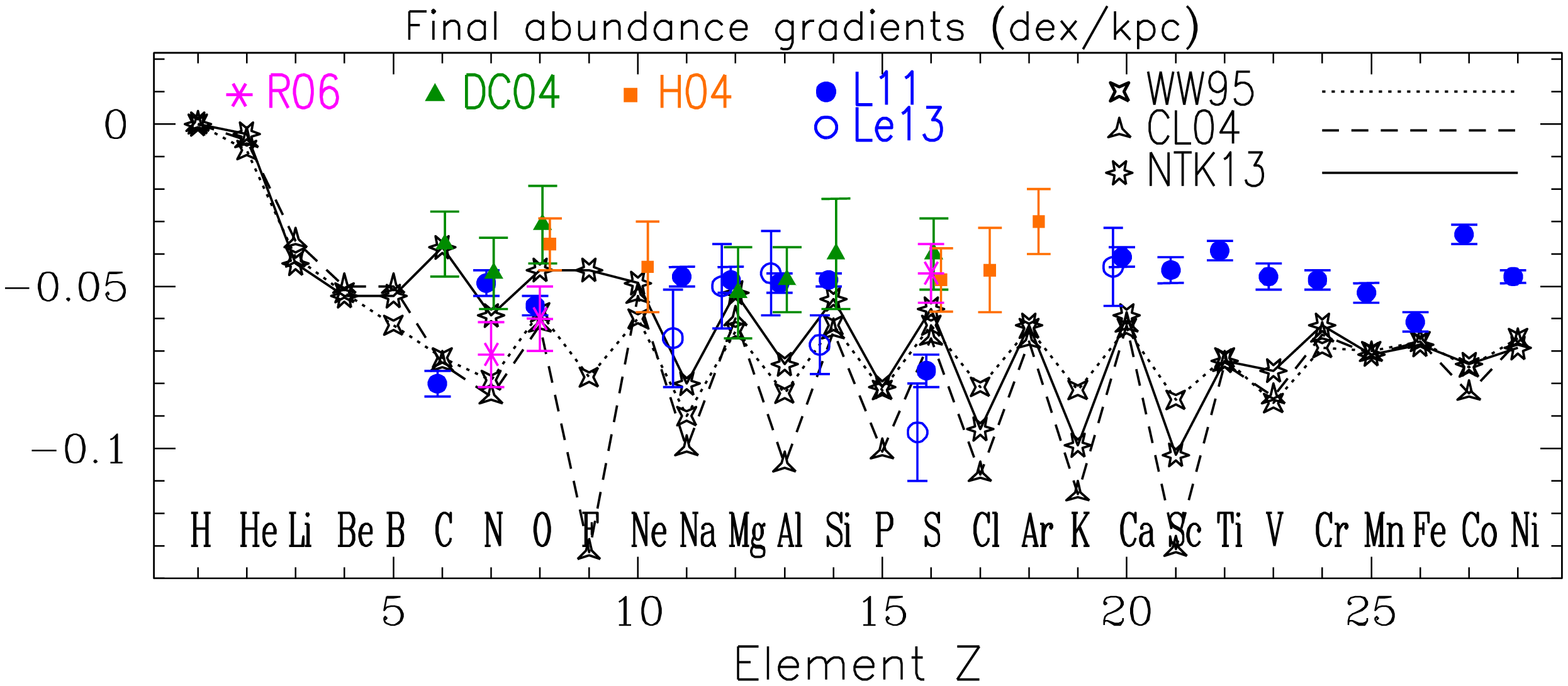}
\caption[]{Present day abundance gradients from H to Ni: models vs. observations. Observations are from H-II regions  \citep{Rudolph2006}, B-stars  \citep{Daflon2004}, Planetary nebulae \citep{Henry2004} and Cepheids  \citep{Luck2011,Luck2011b,Lemasle2013}. Model results
are obtained with the model described here and three different sets of yields, from \cite{Nomoto2013} \cite{WW95}   and \cite{CL2004}, as indicated by the open symbols and discussed in the text.
}
\label{Fig:F_FinalGradients}
\end{center}
\end{figure*}

Concerning the other elements, one sees that observed abundances are systematically higher than solar in the solar neighborhood (except for Mn and Ni), in rough agreement with theoretical expectations.
Our final abundance profiles are globally in agreement with the data, although the obtained slope is, in general, slightly larger than the observationally inferred one, as we discuss below. Finally, the Cepheid data do
not extend into the bar region, as to allow us to probe the predictions of the model there.

In  Fig. \ref{Fig:F_FinalGradients}, we present a quantitative comparison of the gradients of all elements between H and Ni to the data set of Cepheids from \cite{Luck2011}, complemented  with a few other data presented here for completeness. We notice, however, that the abundance profiles are not necessarily perfect exponentials to be fit by straight lines of a single slope in logarithmic space. In our case they are not, and the slope depends on the radial range considered: here we take the range 4-17 kpc, to compare directly with
the slopes provided by \cite{Luck2011} for that same range. The other data sets correspond, however, to different radial ranges, making a direct comparison difficult.

In order to give an idea of the theoretical uncertainties, we provide results for 3 different runs of the same model using 3 different sets of yields. The first one is from \cite{Nomoto2013} (mass range for massive stars: 13-40 \ms) using the low-mass yields of \cite{Karakas2010} (1-6 \ms), as discussed throughout this work. The second set adopts the  massive star yields of \cite{WW95} (12-40 \ms) and the ones of \cite{VandenHoek1997} (0.9-8 \ms) for intermediate mass stars. The third one adopts the massive star yields of \cite{CL2004} (13-35 \ms) and those of  \cite{VandenHoek1997} for intermediate mas stars.
All those yields are metallicity-dependent, but they cover different ranges of masses and metallicities. The limited cover of the mass grid by all the data sets implies the need for  interpolation between the low-mass and massive star ranges,  and extrapolation between the most massive star of the calculated yields and the most massive star of the adopted IMF, here taken to be 100 \ms; these operations  introduces numerical biases in the results. Regarding the metallicities, the yields of \cite{Nomoto2013}  are provided for a finer grid and extend to supersolar metallicities, both features being essential for a 
consistent study of the evolution of the MW disk. The calculations of massive stars involve different ingredients,
e.g. no mass loss in \cite{WW95} and \cite{CL2004} vs mass loss in \cite{Nomoto2013}, different prescriptions for some key nuclear reaction rates or the various mixing mechanisms and the description of the final supernova explosion. Similar differences characterise the physics underlying  the yields of intermediate mass  stars, e.g. the treatment of hot-bottom burning.  Those differences, as well as other important ingredients (not considered in those calculation, like e.g. rotation)  make any attempt for a systematic comparison between sets of yields rather futile. We provide such a comparison only for illustrative purposes, being fully aware that the overall theoretical uncertainties should be   larger than found here. We simply notice that, in all cases we keep the same IMF and the same prescription for the rate and yields
of SNIa. 

%\begin{figure*}
%\begin{center}
%\includegraphics[angle=-90,width=0.99\textwidth]{F_XFe.eps}
%\caption[]{Abundance profiles from C to Zn and comparison to Cepheid data. Model curves correspond to stellar populations of age 8, 4 and 0 Gyr, respectively.
%}
%\label{Fig:F_Other_Cepheids}
%\end{center}
%\end{figure*}

An inspection of Fig. \ref{Fig:F_FinalGradients} shows that the observed gradients of all elements between C and Ni lie in the narrow range of dlogX/dr $\sim$-0.04 to -0.06 \dxpc, with the exception of C and (curiously) S. Taking into account the larger error bars, the data for other tracers are in good agreement with those of Cepheids.

Theoretical results present some features common to all sets of adopted yields.

i) quasi-identical slopes for the Fe-peak elements, dominated by SNIa;

ii) quasi-identical slopes for all $\alpha$-elements beyond Ne;

iii) a distinctive difference in the slopes of even vs odd elements, the former been smaller in absolute value than the latter. To our knowledge, it is the first time that the "odd-even" effect of nucleosynthesis, known to affect the behaviour of the abundance ratios in low metallicity stars (e.g. \citet{Goswami2000}, is put in evidence in the case of the Galactic abundance gradients.

There are also some noticeable differences between the various sets of yields.

- The yields of \cite{Nomoto2013} produce flatter profiles for C,N and O than those of \cite{WW95} or \cite{CL2004}; we find that this is due mainly to the impact of the yields of low-mass stars (from \cite{Karakas2010} in the former case), which 
have a more pronounced "hot-bottom burning" than the yields of \cite{VandenHoek1997}
adopted in the latter two cases.

- the "odd-even" effect appears to be more pronounced with the yields  of \cite{WW95} than with those of \cite{Nomoto2013}, and even more so with the yields of \cite{CL2004}.

- F is a particular case, since \cite{WW95} and \cite{Nomoto2013} include the effect of $\nu$-induced nucleosynthesis during the final SN explosion, while \cite{CL2004} do not.

Comparing model results to observations, one sees a significant offset of the former - by $\sim$0.02 dex downwards - for all elements beyond Cl {\it except Fe}.
For lighter elements, there is satisfactory agreement (within error bars) for the even elements, but significant discrepancy for the odd ones.

Some conclusions may be drawn from the comparison between model and observations on the one hand, and between different sets of yields on the other. First, if the systematic  discrepancy of $\sim$0.01-0.02 dex between model results and observations is confirmed, some key ingredients of the model should be revised: this could be the case, for instance, of the timescales of the infall rate, which should be lower in the outer disk than adopted here (as to favour a more rapid evolution and larger final abundances in that region); alternatively, a metal-enriched composition of the infalling gas could be adopted, instead of the primordial one adopted here.  We find, however, that in the framework of this model the required infall metallicity is $\sim$0.4 \zs, too large for the infalling gas (acceptable metallicities, up to 0.2 \zs, increase the slope by only 0.01 dex/kpc). Second, it is interesting to see whether the abundance gradients present any systematic trends with the atomic number of the element, i.e. either steeper profiles for Fe-peak elements or the "odd-even" effect. If this turns out not to be the case, then
the yields of nucleosynthesis should be revised and the role of the IMF scrutinized (because more massive stars produce larger ratios of $\alpha$/Fe). In any case, the abundance profiles of a large number of elements should be accurately established, on a radial basis as large as possible. Then, the abundance profiles could be used to probe stellar nucleosynthesis, a role played already by abundance ratios in low metallicity halo stars.

 We notice that the study of \cite{Romano2010}, concerning the impact of different sets of yields on the chemical evolution of the Milky Way,
reaches conclusions which present similarities to our own, but also some differences. Their model has no radial migration or radial inflows and uses different prescriptions for the SFR and infall rate than ours; thus, they obtain different abundance profiles, flatter than ours by $\sim$0.02 dex/kpc for C and O (their Fig. 17 left). Still, they adopt the same IMF, while their models 1 and 15 adopt similar sets of yields with our study: the former adopts \cite{WW95} for massive stars and \cite{vandenHoek97} for low and intermediate mass stars (LIMS),
while the latter uses the yields of \cite{Nomoto2013} which include those of \cite{Karakas2010} for LIMS. Thus, some comparison with our results becomes possible. \cite{Romano2010} find that "... for some elements, different choices for the yields provide also different slopes for the gradients; this is the case for carbon and oxygen, while a negligible effect is seen for other elements, including those heavier than Si". We confirm the case for C and O, but we attribute it to the differences in the corresponding yields of LIMS, while \cite{Romano2010} attribute it to the yields of rotating, massive stars with mass loss (we did not study that case here). On the other hand, we confirm that the impact on the slope of the abundance profiles is negligible for other {\it even} elements, but we find a significant effect for the {\it odd} elements, as we emphasized in the previous paragraphs.

\section{Summary}
\label{sec:Summary} 
 
In this work we study the abundance profiles of elements between H and Ni in the MW disk, using a semi-analytical model involving radial motions of gas and stars.
We adopt parametrised descriptions of those radial motions, which are based on N-body simulations for the case of stars and on a simple analytical prescription for the gas radial velocity profile and are inspired by the presence of a bar in the Milky Way.
Other key ingredients of the model is the assumption of a SFR dependent on the molecular gas and the use of a fine grid of recent stellar yields from \cite{Nomoto2013}, which include  up-dated yields of low-mass stars from \cite{Karakas2010} and cover a large range of initial metallicities.
The model reproduces successfully a large number of  observations concerning the solar neighborhood and the disk of the MW, as discussed in KPA14; they include the local age-metallicity relation and metallicity distribution, the $\alpha$/Fe vs Fe/H
relation and the surface density profiles of the thin and thick disks, as well as the profiles of stars, SFR, \hatm \ and \hmol, and the total amounts of gas, stars, SFR, CCSN and SNIa rates in the disk and the bulge of the MW.

In Sec. \ref{sec:Model} we present the key model ingredients and we show how the radial motions affect the profiles of stars, gas, SFR, SNIa and Fe. We find that the effect concerns mainly the inner disk, because of the key role played by the Galactic bar in radial migration according to our assumptions; its effect on the disk is found to be negligible beyond 12 kpc, with the adopted prescriptions.
The effect on the Fe profile in the inner disk is rather small (of the order of 10\% and it is  due to the role of SNIa from long-lived progenitors, which have time enough to migrate away from their birth place. 

In Sec.  \ref{sec:abundance-evol} we present our results and we compare them to a large number of observational data from various metallicity tracers (H-II regions, B-stars, PN, Cepheids and open clusters). We notice that the data base is not homogeneous and does not cover uniformly the radial extent of the MW disk.

Our abundance profiles cannot be characterised by a unique slope, since they flatten progressively towards the outer disk (as a result of the adopted SFR prescription)
and towards the inner disk (as a result of the radial flow induced by the bar).
We find that the abundance profiles flatten with time, as a result of the inside-out formation of the disk. But the observational confirmation of this effect in the MW 
becomes impossible because of the effect of radial migration, which cancels and even inverts it. 
We confirm (Sec. \ref{subsec:Evolution_profiles}) the main effect of radial migration on the abundance profiles found by \cite{Roskar2008}, namely the flattening  of the past abundance profiles of stars, which becomes more pronounced for the older stellar populations. We compare quantitatively our results (Fig. \ref{Fig:Evol_grad}, middle panel) to those of \cite{Minchev2014}, who find similar, albeit somewhat flatter, abundance profiles. 

The evolution of our gaseous abundance profiles is in fair agreement with the extragalactic, high redshift, data compiled by \cite{Jones2013}, which are too uncertain at present, however, to draw firm conclusions (Fig. \ref{Fig:Evol_grad}, bottom panel). In Sec. \ref{sub:OFe} we present the evolution of our [O/Fe] profiles. The evolution of both [Fe/H] and [O/Fe] profiles, modified by radial migration, is encoded in the stellar populations currently present in the local disk and is revealed by preliminary observations of those quantities in stars at different distances from the plane of the disk. Our 1D model lacks the dimension vertical to the plane that would allow us to perform a meaningful comparison to such observations, but the results discussed in Sec. \ref{sub:OFe} are in qualitative agreement with the data (the profiles are expected to flatten with distance from the plane).

In Sec. \ref{sub:Other} we present our results for all elements between C and Ni.
From the theory point of view, we stress the systematic differences in the final abundance profiles due to the different sets of stellar yields (the physics of both low and high mass stars still suffering by several uncertainties). We find a rather good agreement with observationally derived slopes of abundance profiles (assuming they can be described by a single exponential) but also some systematic differences
(see Fig. \ref{Fig:F_FinalGradients}: in particular, we obtain slopes systematically 0.01-0.02 dex larger (in absolute value) than the observed ones. We argue that this difference, if definitively established, could be cured by  some revision of basic ingredients of the model, namely the need for smaller infall timescales in the outer disk or a non-primordial composition for the infalling gas.  We also find an interesting "odd-even" effect of larger slopes for odd elements. This metallicity-dependent effect, already discussed in the context of abundance ratios X/Fe in halo stars, is found here for the first time and it is generic, i.e. it concerns all sets of stellar yields. However, it does not appear in the observational data; if observations are confirmed, then some of the stellar nucleosynthesis results should be revised.

\medskip
\noindent
{\it Acknowledgments}:  We are grateful to an anonymous referee for her/his constructive report.
EA acknowledges financial support to the DAGAL network from the People
Programme  (Marie Curie Actions) of the European Union's Seventh
Framework Programme FP7/2007-2013/ under REA grant agreement number
PITN-GA-2011-289313 and from the CNES (Centre National d'Etudes
Spatiales - France). We also acknowledge partial support from the PNCG
(Programme National Cosmologie et Galaxies - France).

\bibliographystyle{aa} % style aa.bst
\bibliography{Paper4Bib} % your references Yourfile.bib

\end{document}